# Light-induced magnetization precession in GaMnAs

E. Rozkotová, P. Němec[a)], P. Horodyská, D. Sprinzl, F. Trojánek, and P. Malý
*Faculty of Mathematics and Physics, Charles University in Prague, Ke Karlovu 3, 121 16 Prague 2, Czech Republic*

V. Novák, K. Olejník, M. Cukr, and T. Jungwirth
*Institute of Physics ASCR v.v.i., Cukrovarnická 10, 162 53 Prague, Czech Republic*

We report dynamics of the transient polar Kerr rotation (KR) and of the transient reflectivity induced by femtosecond laser pulses in ferromagnetic (Ga,Mn)As with no external magnetic field applied. It is shown that the measured KR signal consist of several different contributions, among which only the oscillatory signal is directly connected with the ferromagnetic order in (Ga,Mn)As. The origin of the light-induced magnetization precession is discussed and the magnetization precession damping (Gilbert damping) is found to be strongly influenced by annealing of the sample.

(Ga,Mn)As is the most intensively studied member of the family of diluted magnetic semiconductors with carrier-mediated ferromagnetism [1]. The sensitivity of ferromagnetism to concentration of charge carriers opens up the possibility of magnetization manipulation on the picosecond time scale using light pulses from ultrafast lasers [2]. Photoexcitation of a magnetic system can strongly disturb the equilibrium between the mobile carriers (holes), localized spins (Mn ions), and the lattice. This in turn triggers a variety of dynamical processes whose characteristic time scales and strengths can be investigated by the methods of time-resolved laser spectroscopy [2]. In particular, the magnetization reversal dynamics in various magnetic materials attracts a significant attention because it is directly related to the speed of data storage in the magnetic recording [3]. The laser-induced precession of magnetization in ferromagnetic (Ga,Mn)As has been recently reported by two research groups [4-6] but the physical processes responsible for it are still not well understood. In this paper we report on simultaneous measurements of the light-induced magnetization precession dynamics and of the dynamics of photoinjected carriers.

The experiments were performed on a 500 nm thick ferromagnetic $Ga_{1-x}Mn_xAs$ film with $x = 0.06$ grown by the low temperature molecular beam epitaxy (LT-MBE) on a GaAs(001) substrate. We studied both the as-grown sample, with the Curie temperature $T_C \approx 60$ K and the conductivity of 120 $\Omega^{-1}cm^{-1}$, and the sample annealed at 200°C for 30 hours, with $T_C \approx 90$ K and the conductivity of 190 $\Omega^{-1}cm^{-1}$; using the mobility vs. hole density dependence typical for GaMnAs [7] we can roughly estimate their hole densities as $1.5 \times 10^{20}$ $cm^{-3}$ and $3.4 \times 10^{20}$ $cm^{-3}$, respectively. Magnetic properties of the samples were measured using a superconducting quantum interference device (SQUID) with magnetic field of 20 Oe applied along different crystallographic directions. The photoinduced magnetization dynamics was studied by the time-resolved Kerr rotation (KR) technique [2] using a femtosecond titanium sapphire laser (Tsunami, Spectra Physics). Laser pulses, with the time width of 80 fs and the repetition rate of 82 MHz, were tuned to 1.54 eV. The energy fluence of the pump pulses was typically 15 µJ.cm$^{-2}$ and the probe pulses were always at least 10 times weaker. The polarization of the pump pulses was either circular or linear, while the probe pulses were

---

[a)] Electronic mail: nemec@karlov.mff.cuni.cz

linearly polarized (typically along the [010] crystallographic direction in the sample, but similar results were obtained also for other orientations). The rotation angle of the polarization plane of the reflected probe pulses was obtained by taking the *difference* of signals measured by detectors in an optical bridge detection system [2]. Simultaneously, we measured also the *sum* of signals from the detectors, which corresponded to a probe intensity change due to the pump induced modification of the sample reflectivity. The experiment was performed with no external magnetic field applied. However, the sample was cooled in some cases with no external magnetic field applied or alternatively with a magnetic field of 170 Oe applied along the [-110] direction.

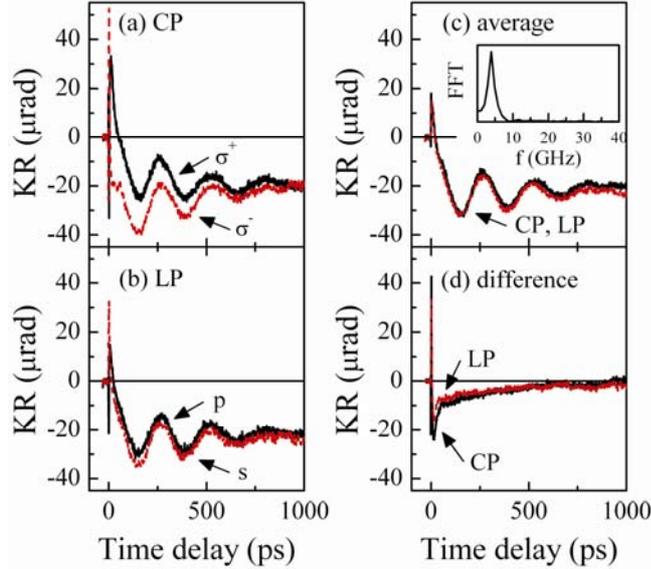

Fig. 1. Dynamics of photoinduced Kerr rotation angle (KR) measured for the as-grown sample at 10 K. (a) KR measured for $\sigma^+$ and $\sigma^-$ circularly polarized (CP) pump pulses; (b) KR measured for *p* and *s* linearly polarized (LP) pump pulses. Polarization-independent part (c) (polarization-dependent part (d)) of KR signal, which was computed from the measured traces as an average of the signals (a half of the difference between the signals) detected for pump pulses with the opposite CP (LP). Inset: Fourier transform of the oscillations. No external magnetic field was applied during the sample cooling.

In Fig. 1 we show typical temporal traces of the transient angles of KR measured for the as-grown sample at 10 K. The KR signal was dependent on the light polarization but there were certain features present for both the circular (Fig. 1 (a)) and linear (Fig. 1 (b)) polarizations. In Fig. 1 (c) we show the polarization-independent part of the measured KR signal, which was the same for circular and linear polarization of pump pulses. On the other hand, the amplitude of the polarization-dependent part of the signal (Fig. 1 (d)) was larger for the circular polarization. The interpretation of the polarization-dependent part of the signal is significantly complicated by the fact that the circularly polarized light generates spin-polarized carriers (electrons in particular), whose contribution to the measured KR signal can even exceed that of ferromagnetically coupled Mn spins [8]. In the following we concentrate on the polarization-independent part of the KR signal (Fig. 1 (c)). This signal can be fitted well (see Fig. 2) by an exponentially damped sine harmonic oscillation superimposed on a pulse-like function:

$$KR(t) = A\exp(-t/\tau_D)\sin(\omega t + \varphi) + B\left[1 - \exp(-t/\tau_1)\right]\exp(-t/\tau_2). \tag{1}$$

The oscillatory part of the KR signal is characterized by the amplitude ($A$), damping time ($\tau_D$), angular frequency ($\omega = 2\pi f$), and phase ($\varphi$). The pulse-like part of the KR signal is described by the amplitude ($B$), rise time ($\tau_1$), and decay time ($\tau_2$). In the inset of Fig. 2 we show the dynamics of the sample reflectivity change $\Delta R/R$. This signal monitored the change of the complex index of refraction of the sample due to carriers photoinjected by the pump pulse. From the dynamics of $\Delta R/R$ we can conclude that the population of photogenerated free carriers (electrons in particular [9]) decays within ≈ 50 ps after the photoinjection. This rather short lifetime of free electrons is similar to that reported for the low temperature grown GaAs (LT-GaAs), which is generally interpreted as a consequence of a high concentration of nonradiative recombination centers induced by the low temperature growth mode of the MBE [9]. It is also clearly apparent from the inset of Fig. 2 that the KR data can be fitted well by Eq. (1) only for time delays larger than ≈ 50 ps (i.e., just after the population of photoinjected free electrons nonradiatively decayed). We will come back to this point later.

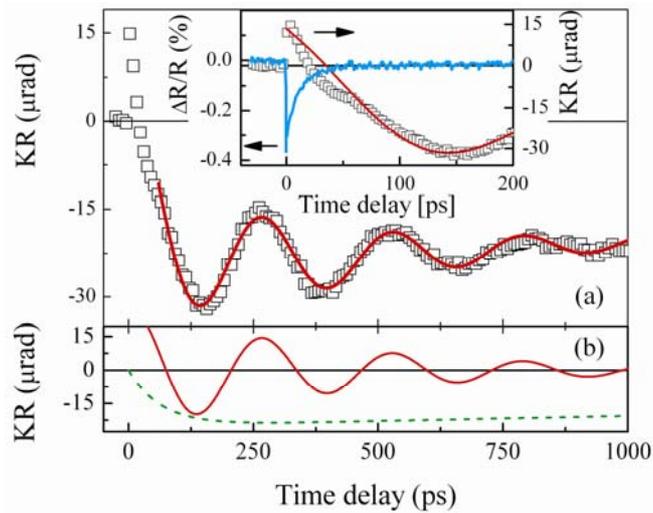

Fig. 2. The fitting procedure applied to the polarization-independent part of KR signal. (a) The measured data from Fig. 1 (c) (points) are fitted (solid line) by a sum of the exponentially damped sine harmonic oscillation (solid line in part (b)) and the pulse-like KR signal (dashed line in part (b)). Inset: Dynamics of the reflectivity change (thick solid line) and the detail of the fitted KR signal.

In Fig. 3(a) we show the intensity dependence of $A$ and $B$, and in Fig. 3(b) of $\omega$ and $\tau_D$ measured at 10 K. For the increasing intensity of pump pulses the magnitudes of $A$ and $B$ were increasing, $\omega$ was decreasing and the values of $\tau_D$ were not changing significantly. The application of magnetic field applied along the [-110] direction during the sample cooling modified the value of $\omega$. For 10 K (and pump intensity $I_0$) the frequency decreased from 24.5 to 20 GHz (open and solid point in Fig. 3 (d), respectively). The measured temperature dependence of $A$ and $B$ (Fig. 3 (c)) revealed that the oscillatory signal vanished above $T_C$, while a certain fraction of the pulse-like KR signal persisted even above $T_C$. This shows that only the oscillatory part of the KR signal was directly connected with the *ferromagnetic* order in (Ga,Mn)As. (It is worth noting that also the polarization-dependent part of the KR signal was non-zero even above $T_C$.) The frequency of oscillations was decreasing with the sample temperature (Fig. 3 (d)), but the values of $\tau_D$ were not changing significantly (not shown here).

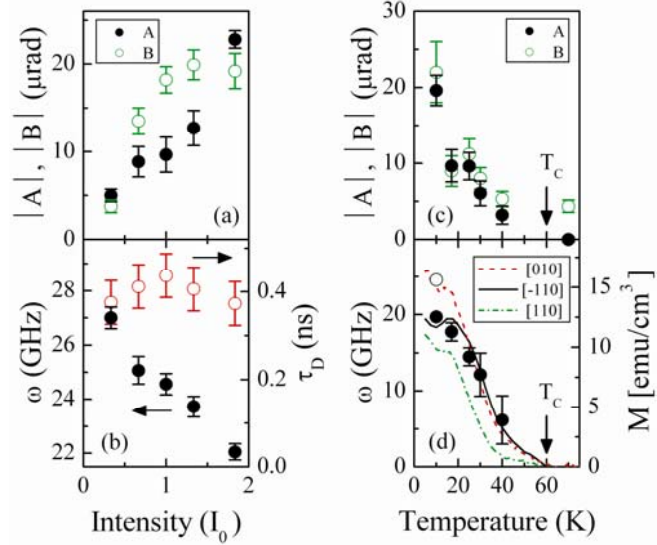

Fig. 3. Intensity dependence of $|A|$ and $|B|$ (a), $\omega$ and $\tau_D$ (b) measured at 10 K; $I_0 = 15$ μJ.cm$^{-2}$, no external magnetic field was applied during the sample cooling. (c), (d) Temperature dependence of $|A|$, $|B|$ and $\omega$ (points) measured at pump intensity $I_0$. The open point in (d) was obtained for the sample cooled with no external magnetic field applied and the data in (c) and the solid points in (d) were obtained for the sample cooled with magnetic field applied along the [-110]. The lines in (d) are the temperature dependence of the sample magnetization projections to different crystallographic directions measured by SQUID.

The photoinduced magnetization precession was reported by A. Oiwa *et al.*, who attributed it to the precession of ferromagnetically coupled Mn spins induced by a change in magnetic anisotropy initiated by an increase in hole concentration [4]. It was also shown that the photoinduced magnetization precession and the ferromagnetic resonance (FMR) can provide similar information [4]. Magnetic anisotropy in (Ga,Mn)As is influenced by the intrinsic cubic anisotropy, which is arising from its zinc-blende symmetry, and by the uniaxial anisotropy, which is a result of a strain induced by different lattice constants of GaMnAs and the substrate. For the standard stressed GaMnAs films with Mn content above 2% grown on GaAs substrates the magnetic easy axes are in-plain. Consequently, the measured polar Kerr rotation is not sensitive to the steady state magnetization of the sample, but only to the light-induced transient out-of-plane magnetization due to the polar Kerr effect [2]. In our experiment, the pump pulses with a fluence $I_0 = 15$ μJ.cm$^{-2}$ photoinjected electron-hole pairs with an estimated concentration $\Delta p = \Delta n \approx 8 \times 10^{17}$ cm$^{-3}$. This corresponded to $\Delta p/p \approx 0.5\%$ and such a small increase in the hole concentration is highly improbable to lead to any sizable change of the sample anisotropy [1]. Another hypothesis about the origin of the light-induced magnetization precession was reported recently by J. Qi *et al.* [6]. The authors suggested that not only the transient increase in local hole concentration $\Delta p$ but also the local temperature increase $\Delta T$ contributes to the change of anisotropy constants. This modification of the sample anisotropy changes in turn the direction of the in-plane magnetic easy axis and, consequently, triggers a precessional motion of the magnetization around the altered magnetic anisotropy field $H_{anis}^{Mn}$. The magnitude of $H_{anis}^{Mn}$ decreases as $T$ (the sample temperature) or $\Delta T$ increases, primarily due to the decrease in the cubic anisotropy constant $K_{1c}$ [6]. Our samples exhibit in-plane easy axis behavior typical for stressed GaMnAs layers grown on GaAs substrates. To characterize their in-plane anisotropy we measured the temperature dependent magnetization projections to [110], [010], and [-110] crystallographic directions – the results are shown in Fig. 3 (d) and in inset of Fig.4 for the as-grown and the annealed sample, respectively. At low temperatures the cubic anisotropy dominates (as indicated by the maximal projection measured along the [010] direction) but the uniaxial in-plane component is not negligible and the sample magnetization is slightly tilted from the [010] direction towards the [-110] direction. Both samples exhibit rotation of magnetization direction in the temperature region 10-25K, which is in agreement with the expected fast weakening of the cubic component with an increasing temperature. In our experiment, the excitation fluence $I_0$ led to $\Delta T \approx 10$ K (as estimated from the GaAs specific heat of 1 mJ/g/K [6]) that can be sufficient for a change of the easy axis position. This temperature-based hypothesis about the origin of magnetization precession is supported also by our observation that the oscillations were not fully developed *immediately* after the photoinjection of carriers but only after $\approx 50$ ps when phonons were emitted by the nonradiative decay of the population of free electrons (see inset in Fig. 2). We also point out that the measured precession frequency $\omega$ and the sample magnetization $M$ (measured by SQUID) had very similar temperature dependence (see Fig. 3(d)).

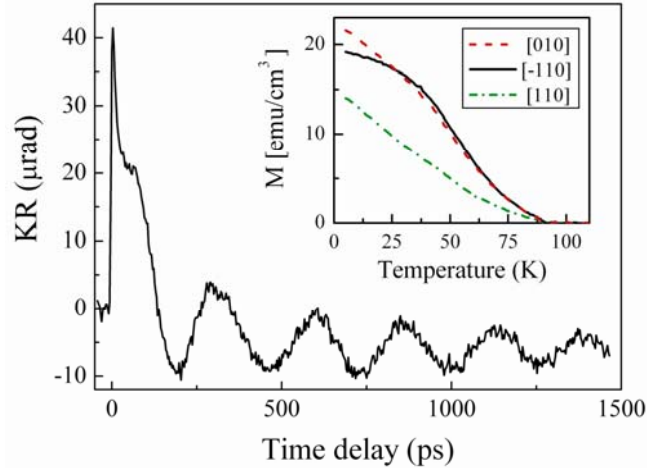

Fig. 4. Polarization-independent part of KR signal measured for the annealed sample at 10 K; $I_0 = 15$ μJ.cm$^{-2}$, the sample was cooled with magnetic field applied along the [-110]. Inset: Temperature dependence of the sample magnetization projections to different crystallographic directions measured by SQUID.

An example of the results measured for the annealed sample is shown in Fig. 4. The analysis of the data revealed that at similar conditions the precession frequency was slightly higher in the annealed sample (20 GHz and 24 GHz for the as-grown and the annealed sample, respectively). However, a major effect of the sample annealing was on the oscillation damping time $\tau_D$, which increased from 0.4 ns to 1.1 ns. This prolongation of $\tau_D$ can be attributed to the improved quality of the annealed sample, which is indicated by the higher value of $T_C$ and by the more Brillouin-like temperature dependence of the magnetization (cf. Fig. 3 (d) and inset in Fig. 4). The damping of oscillations is connected with the precession damping in the Landau-Lifshitz-Gilbert equation [1]. The exact determination of the intrinsic Gilbert damping coefficient $\alpha$ from the measured data is not straightforward because it is difficult to decouple the contribution due to the inhomogeneous broadening [10]. In Ref. 6 the values of $\alpha$ from 0.12 to 0.21 were deduced for the as-grown sample from the analysis of the oscillatory KR signal. The time-domain KR should provide similar information as the frequency-domain based FMR, where the relaxation rate of the magnetization is connected with the peak-to-peak ferromagnetic resonance linewidth $\Delta H_{pp}$ [10]. Indeed both methods showed that the relaxation rate of the magnetization is considerably slower in the annealed samples (as indicated by the prolongation of $\tau_D$ in our experiment and by the reduction of $\Delta H_{pp}$ in FMR [10]).

In conclusion, we studied the transient Kerr rotation (KR) and the reflectivity change induced by laser pulses in (Ga,Mn)As with no external magnetic field applied. We revealed that the measured KR signals consisted of several different contributions and we showed that only the oscillatory KR signal was directly connected with the ferromagnetic order in (Ga,Mn)As. Our data indicated that the phonons emitted by photoinjected carriers during their nonradiative recombination in (Ga,Mn)As can be responsible for the magnetic anisotropy change that was triggering the magnetization precession. We also observed that the precession damping was strongly suppressed in the annealed sample, which reflected its improved magnetic properties. This work was supported by Ministry of Education of the Czech Republic in the framework of the research centre LC510, the research plans MSM0021620834 and AV0Z1010052, by the Grant Agency of the Charles University in Prague under Grant No. 252445, and by the Grant Agency of Academy of Sciences of the Czech Republic Grants FON/06/E001, FON/06/E002, and KAN400100652.